\documentclass{PoS}
\usepackage{amsmath,bm}
\usepackage{subfigure}
\usepackage{graphicx}
\usepackage{epsfig}
\title{The production of $b\bar{b}$ dijet in heavy-ion collisions at the LHC}

\ShortTitle{The production of $b\bar{b}$ dijet in heavy-ion collisions at the LHC}

\author{\speaker{Sa Wang}$^a$, Wei Dai$^b$ , Shan-Liang Zhang$^a$ , Ben-Wei Zhang$^a$ , Enke Wang$^a$\\
        \llap{$^a$} Key Laboratory of Quark \& Lepton Physics (MOE) and Institute of Particle Physics, Central China Normal University, Wuhan 430079, China\\
        \llap{$^b$}School of Mathematics and Physics, China University
of Geosciences, Wuhan 430074, China\\
        E-mail: \email{wangsa@mails.ccnu.edu.cn},  \\
      \qquad\qquad\email{weidai@cug.edu.cn},  \\
    \qquad\qquad\email{bwzhang@mail.ccnu.edu.cn}}

\abstract{We report our recent theoretical calculations for $b\bar{b}$ dijet production in high-energy nuclear collisions.
The NLO+parton shower (PS) event generator SHERPA has been employed to provide the pp baseline of $b\bar{b}$ dijet production. A framework which combines Langevin transport equation to describe the evolution of heavy quarks  and the higher-twist scheme to consider the inelastic energy loss of both light and heavy partons has been implemented. Within this framework, we present the theoretical result of transverse momentum imbalance $x_J=p_{T1}/p_{T2}$ both for inclusive dijet and $b\bar{b}$ dijet in Pb+Pb collision at 5.02 TeV. The energy loss of b jet would shift $x_J$ to smaller value relative to p+p reference which is consistent with the CMS data. In addition, we show the medium modification for angular correlation of $b\bar{b}$ dijet in A+A collision at $\sqrt{s_{NN}}=5.02$~TeV. We observe a stronger suppression in small $\Delta \phi=|\phi_{b1}-\phi_{b2}|$ region where the gluon splitting processes dominate relative to large $\Delta \phi$ region. The difference leads to a modest suppression on near side ($\Delta \phi\sim 0$) and enhancement on away side ($\Delta\phi\sim\pi$). }

\FullConference{International Conference on Hard and Electromagnetic Probes of High-Energy Nuclear Collisions\\
		30 September - 5 October 2018\\
		Aix-Les-Bains, Savoie, France}

\begin{document}

\section{Introduction}
\label{Introduction}
The ``jet quenching'' effect has been proposed as a good probe to study the properties of the quark-gluon plasma (QGP) in ultrarelativistic heavy-ion collision (HIC)~\cite{Wang:1991xy}. Because of the large mass of b quark, the $b\bar{b}$ dijet is a nice channel to test the flavour dependence of jet quenching. The heavy flavour production could be categorized into three mechanisms~\cite{Norrbin:2000zc}: flavour creation (FCR), flavour excitation (FEX) and gluon splitting (GSP). To describe the $b\bar{b}$ dijet in p+p collision succesfully, especially in considering its azimuthal angular correlation, the next-to-leading order (NLO) matched parton shower (PS) event generator is required for Monte Carlo simulation~\cite{Dai:2018mhw,Zhang:2018urd}. Furthermore, to study the observables of heavy-flavoured jets, it's still a challenge to simultaneously describe both heavy and light partons inside the jets in the same framework. In this talk, we will give our latest results for the $p_T$ imbalance and angular correlation of the $b\bar{b}$ dijet~\cite{Dai:2018mhw} in p+p and Pb+Pb collisions as well as the comparison with the recent CMS data~\cite{Sirunyan:2018jju}.

\section{p+p baseline}
\label{sec:pp baseline}
\par
\begin{figure}[]
\center
 \subfigure[]{\label{fig:s7000}
  \epsfig{file=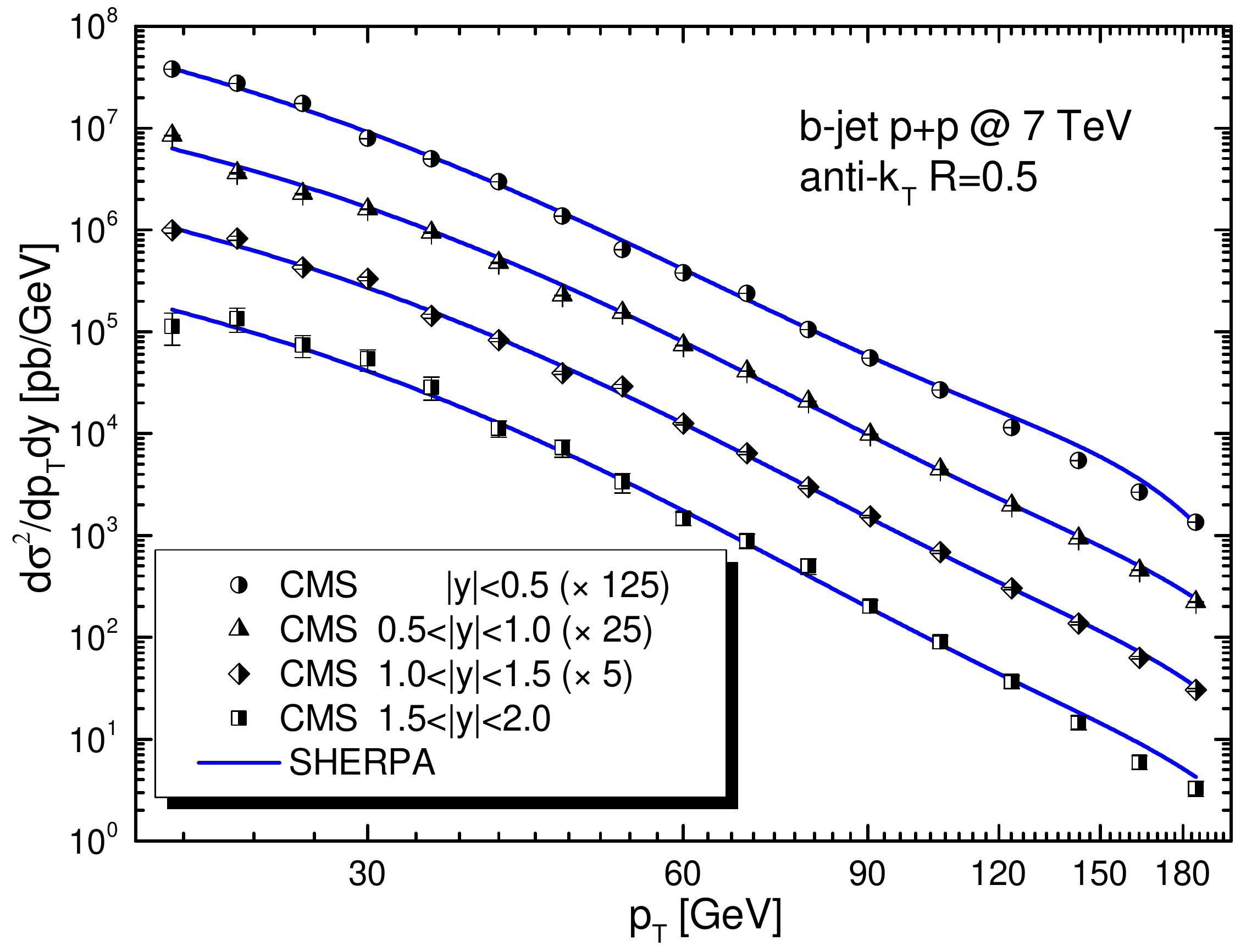, width=0.4\textwidth, clip=}}
 \subfigure[]{\label{fig:phi7000}
  \epsfig{file=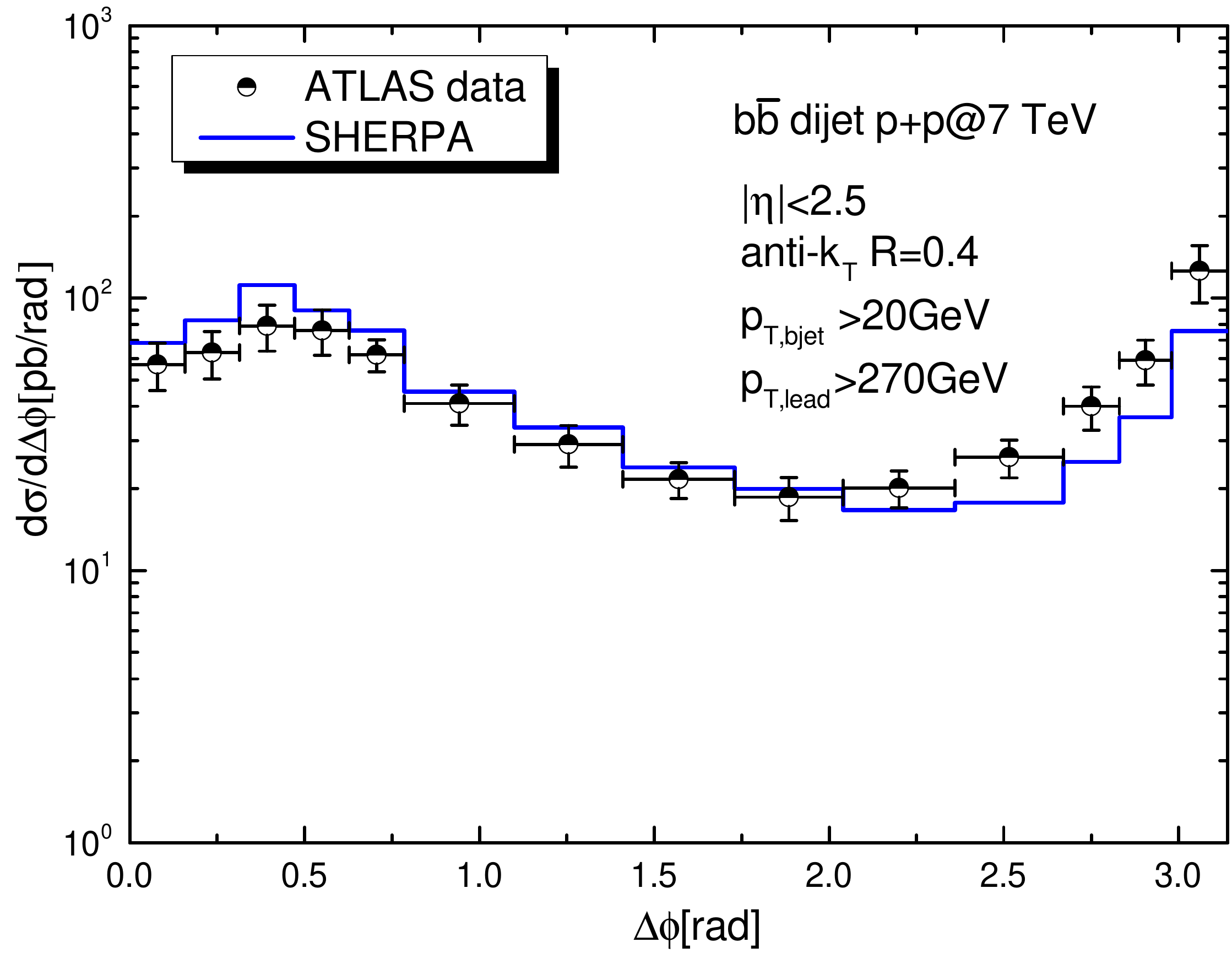, width=0.4\textwidth, clip=}}
  \caption{\scriptsize{(a): Differential cross section of inclusive b-jet production as a function of jet transverse momentum at 7 TeV provided by SHERPA. (b): Differential cross section of $b\bar{b}$ dijet production as a function of $\Delta \phi=|\phi_{b1}-\phi_{b2}|$ at 7 TeV provided by SHERPA.}}
  \label{fig:baseline}
\end{figure}

 In our work, the NLO+PS Monte Carlo event generator SHERPA~\cite{Gleisberg:2008ta} has been employed to provide the pp baseline for the productions of inclusive b-jet and $b\bar{b}$ dijet. The NLO parton distribution functions with 5-flavour scheme sets~\cite{Ball:2014uwa} have been chosen. FASTJET~\cite{Cacciari:2011ma} with anti-$k_{\rm T}$ algorithm has been used for jet reconstruction. We find that SHERPA could provide a good description for the experiment data in p+p collision measured by CMS~\cite{Chatrchyan:2012dk} and ATLAS~\cite{Aaboud:2016jed} collaboration, as shown in Fig.~\ref{fig:baseline}.

\section{Framework of in-medium parton energy loss}
\label{sec:evolution}
To describe the propagation and energy loss of heavy quarks in QGP, in our simulation, the modified discrete Langevin equations have been used~\cite{Zhou:2016vwq},
\begin{align*}
\vec{x}(t+\Delta t)&=\vec{x}(t)+\frac{\vec{p}(t)}{E}\Delta t \tag{1}\\
\vec{p}(t+\Delta t)&=\vec{p}(t)-\Gamma\vec{p} \Delta t+\vec{\xi}(t)-\vec{p}_g\tag{2}
\end{align*}
 According to the fluctuation-dissipation theorem, the relationship between the drag coefficient $\Gamma$ and  diffusion coefficient $\kappa$ could be expressed as $\kappa=2ET\Gamma=\frac{2T^2}{D_s}$, where $D_s$ denoting spacial diffusion coefficient has been fixed at $2\pi TD_s=4$ from the Lattice QCD calculation. The Hard-Thermal Loop result for the elastic energy loss of light partons~\cite{Neufeld:2010xi} has also been considered. The last term in Eq.(2) represents the modification because of the medium-induced gluon radiation based on the higher-twist scheme~\cite{Guo:2000nz,Zhang:2003yn,Zhang:2003wk,Majumder:2009ge,Cao:2017hhk}:
\begin{align*}
\frac{dN}{dxdk^{2}_{\perp}dt}=\frac{2\alpha_{s}C_sP(x)\hat{q}}{\pi k^{4}_{\perp}}\sin^2(\frac{t-t_i}{2\tau_f})(\frac{k^2_{\perp}}{k^2_{\perp}+x^2M^2})^4\tag{3}
\end{align*} 

\section{Transverse momentum imbalance and angular correlation  }
\label{sec:results}

\par
\begin{figure}[]
\center
 \subfigure{\label{fig:xjb0-10}
  \epsfig{file=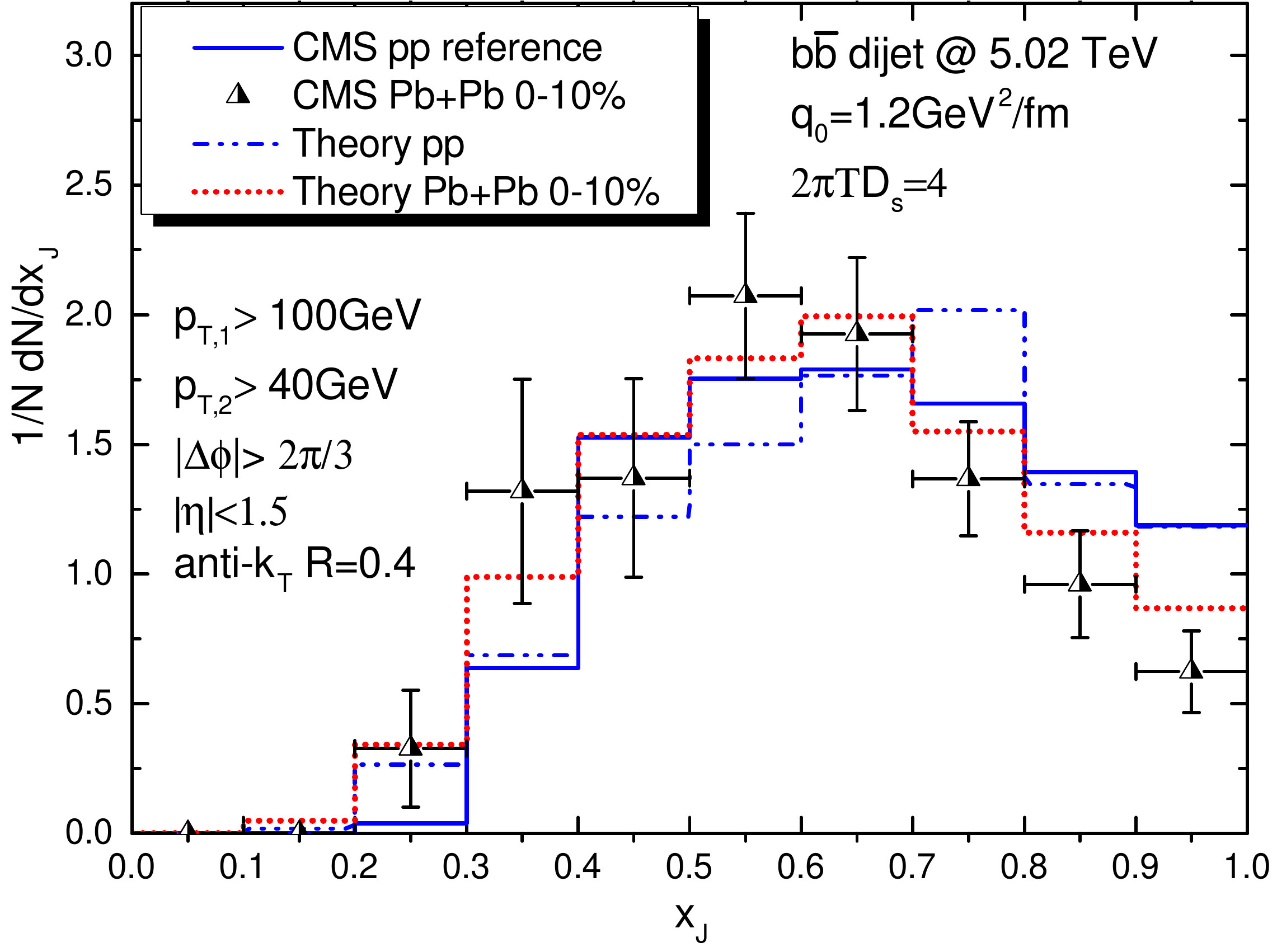, width=0.4\textwidth}}
 \subfigure{\label{fig:xjb30-100}
  \epsfig{file=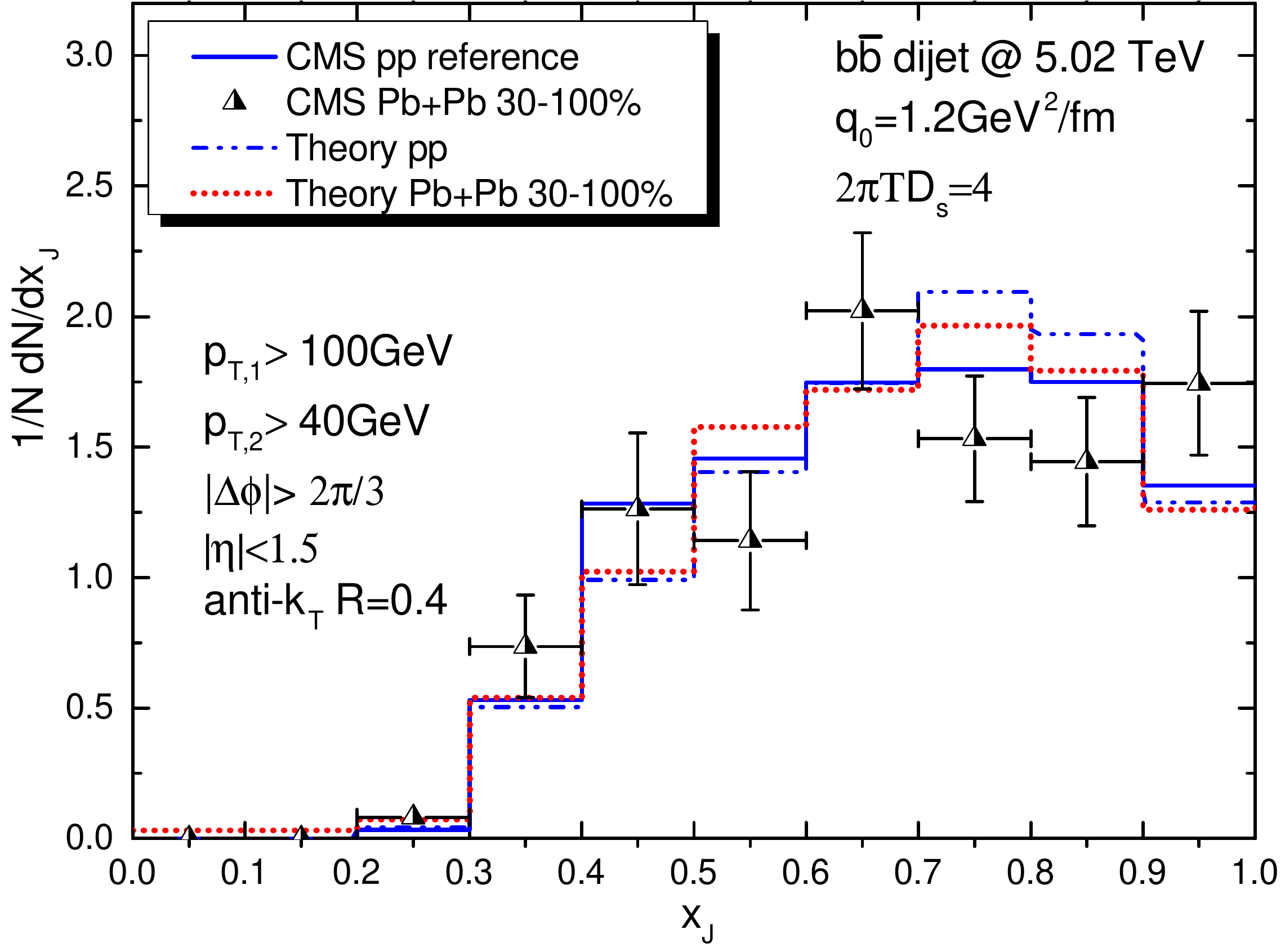,width=0.4\textwidth}}
  \caption{\scriptsize{Normalized $x_J$ distribution of $b\bar{b}$ dijets in (left)~$0-10\%$, (right)~$30-100\%$ Pb+Pb collisions at $\sqrt{s_{NN}}=5.02~TeV$ compared with the experimental data~\cite{Sirunyan:2018jju} and their pp references respectively.}}
  \label{fig:xjb5020}
\end{figure}

 We show the theoretical results for the medium modification of transverse momentum imbalance of $b\bar{b}$ dijet in central  ($0-10\%$) and periperal ($30-100\%$) Pb+Pb collision at 5.02~TeV and compare them with the recent CMS data~\cite{Sirunyan:2018jju} in Fig.~\ref{fig:xjb5020}. Relative to the pp reference, the jet quenching effect would shift the normalized $x_J$ distribution to smaller value which is consistent with the experiment data. And the shift is not visible in the peripheral Pb+Pb collision since much smaller energy loss suffered on the $b\bar{b}$ dijet.


\begin{figure}[]
\center
 \subfigure{\label{fig:avexjdijet}
  \epsfig{file=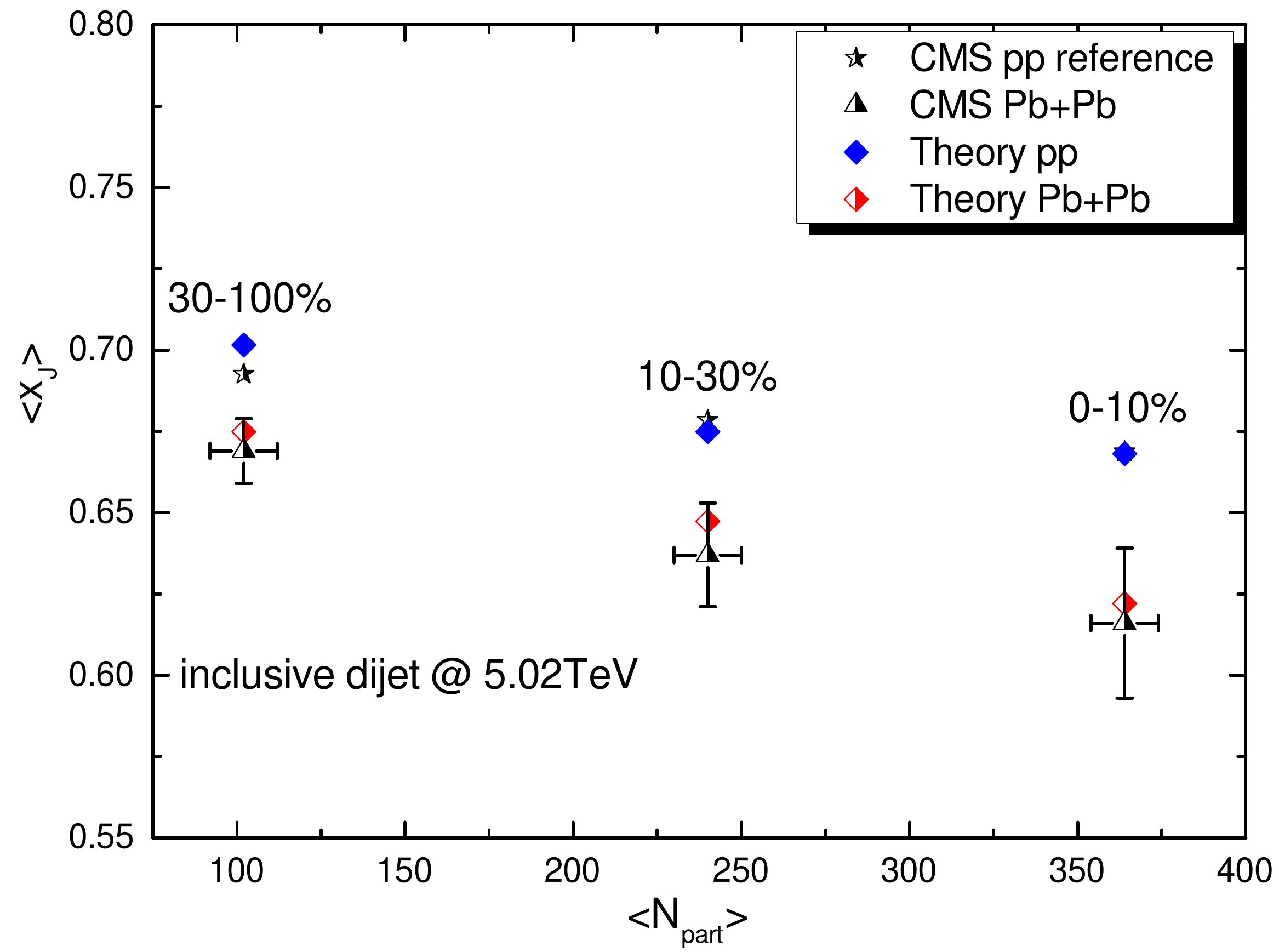, width=0.4\textwidth, clip=}}
 \subfigure{\label{fig:avexjbjet}
  \epsfig{file=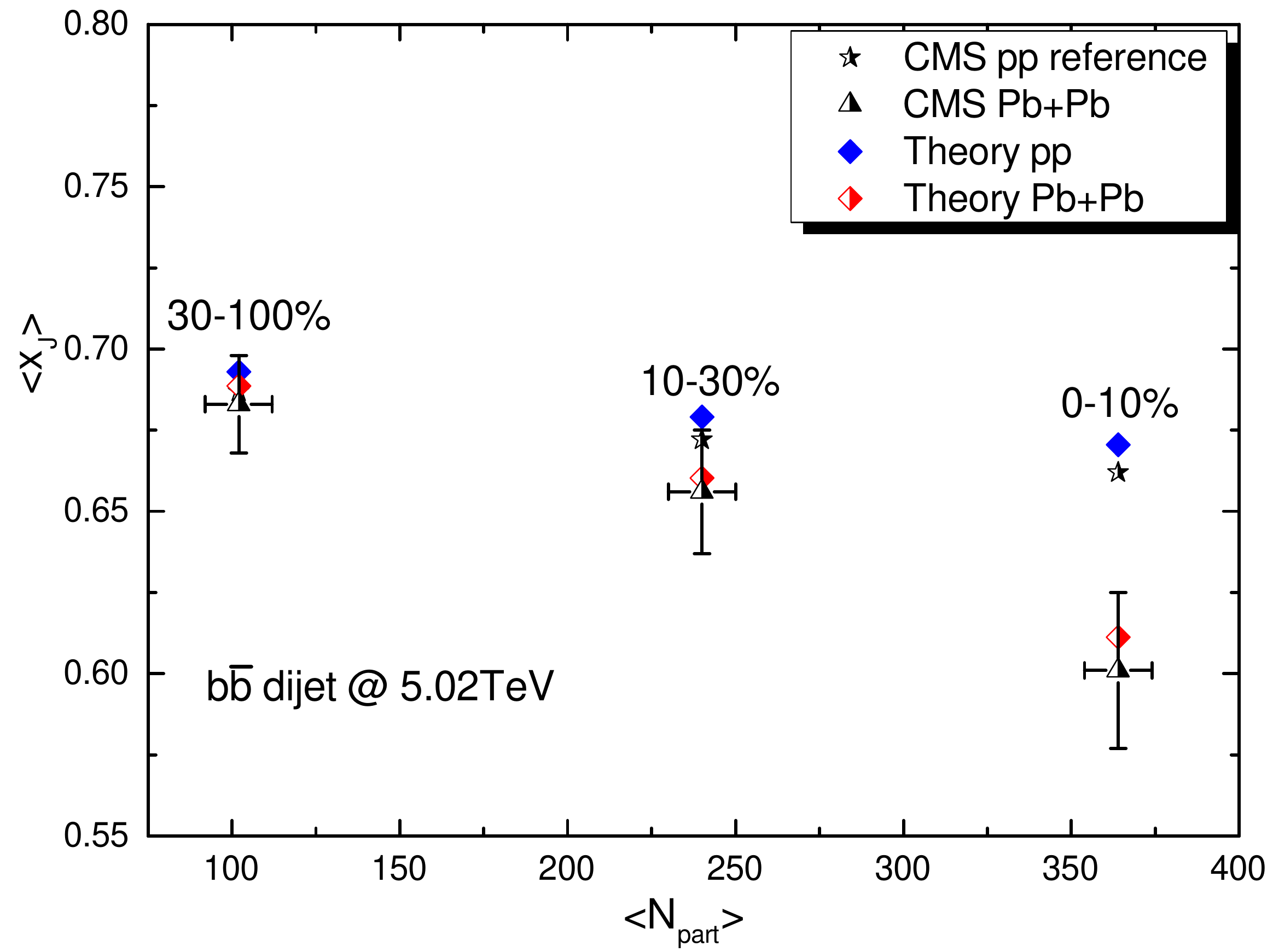, width=0.4\textwidth, clip=}}
  \caption{\scriptsize{Simulated averaged $x_J$ of inclusive dijet (left) and $b\bar{b}$ dijet (right) as a function of number of participant in p+p and Pb+Pb collision compared with the p+p reference and experimental data~\cite{Sirunyan:2018jju} in Pb+Pb collisions respectively .}}
  \label{fig:avexj}
\end{figure}

We also plot the averaged $x_J$ distribution as a function of number of participant both for inclusive dijet and $b\bar{b}$ dijet in Pb+Pb collision in Fig.~\ref{fig:avexj}. Even the smearing treatment decreases the $\left\langle x_J \right\rangle$   with the increasing centrality, the reduction of $\left\langle x_J \right\rangle$ caused by jet in-medium energy loss is much visible for  both inclusive dijet and $b\bar{b}$ dijet in the central Pb+Pb collision. And we also notice that, in the peripheral Pb+Pb collision, the reduction of $b\bar{b}$ dijet $\left\langle x_J\right\rangle$ relative to its pp reference is smaller than that the case in the inclusive dijet.

\begin{figure}[]
\center
 \subfigure[]{\label{fig:phi5020}
  \epsfig{file=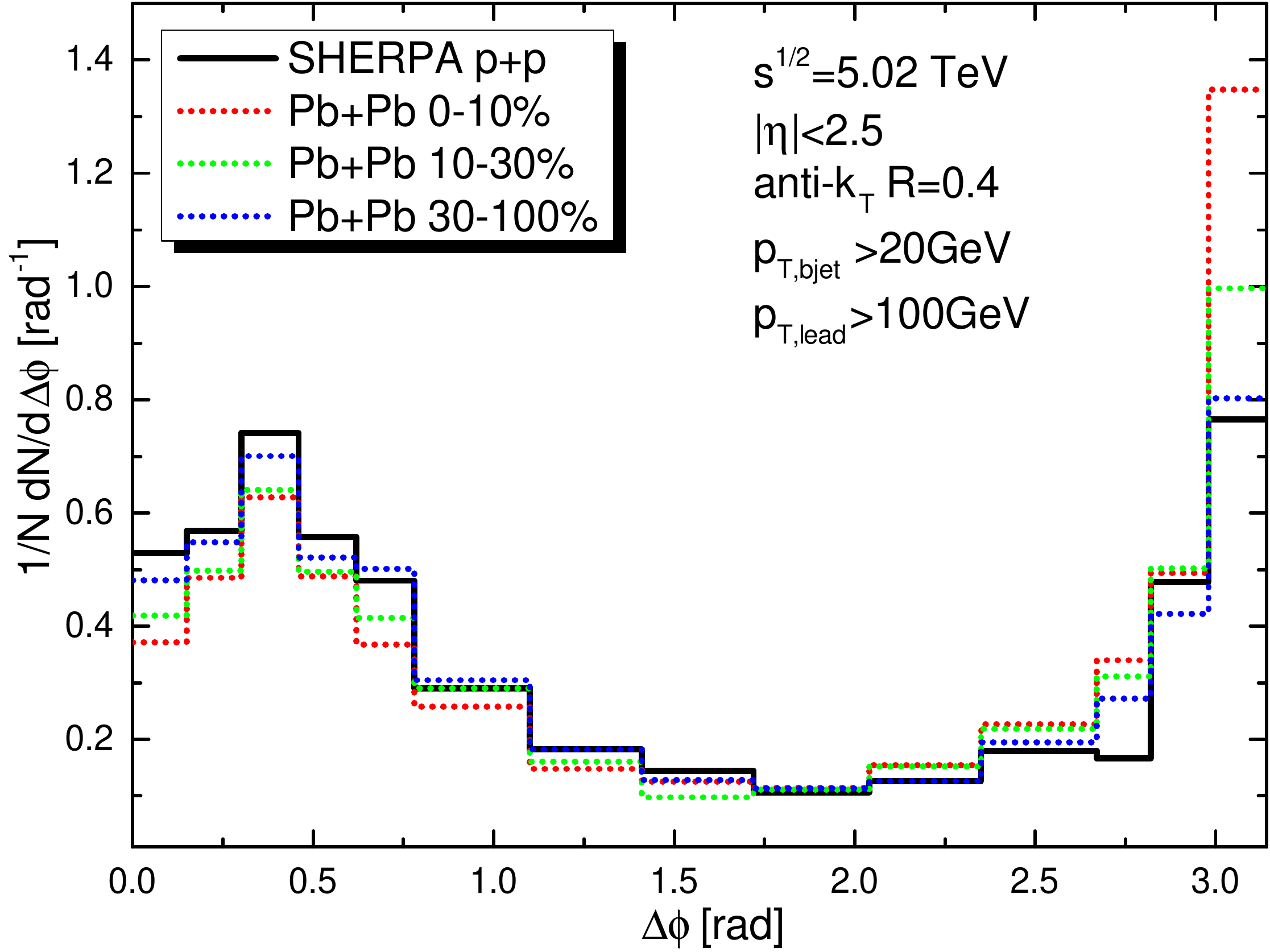, width=0.3\textwidth, clip=}}
 \subfigure[]{\label{fig:R5020}
  \epsfig{file=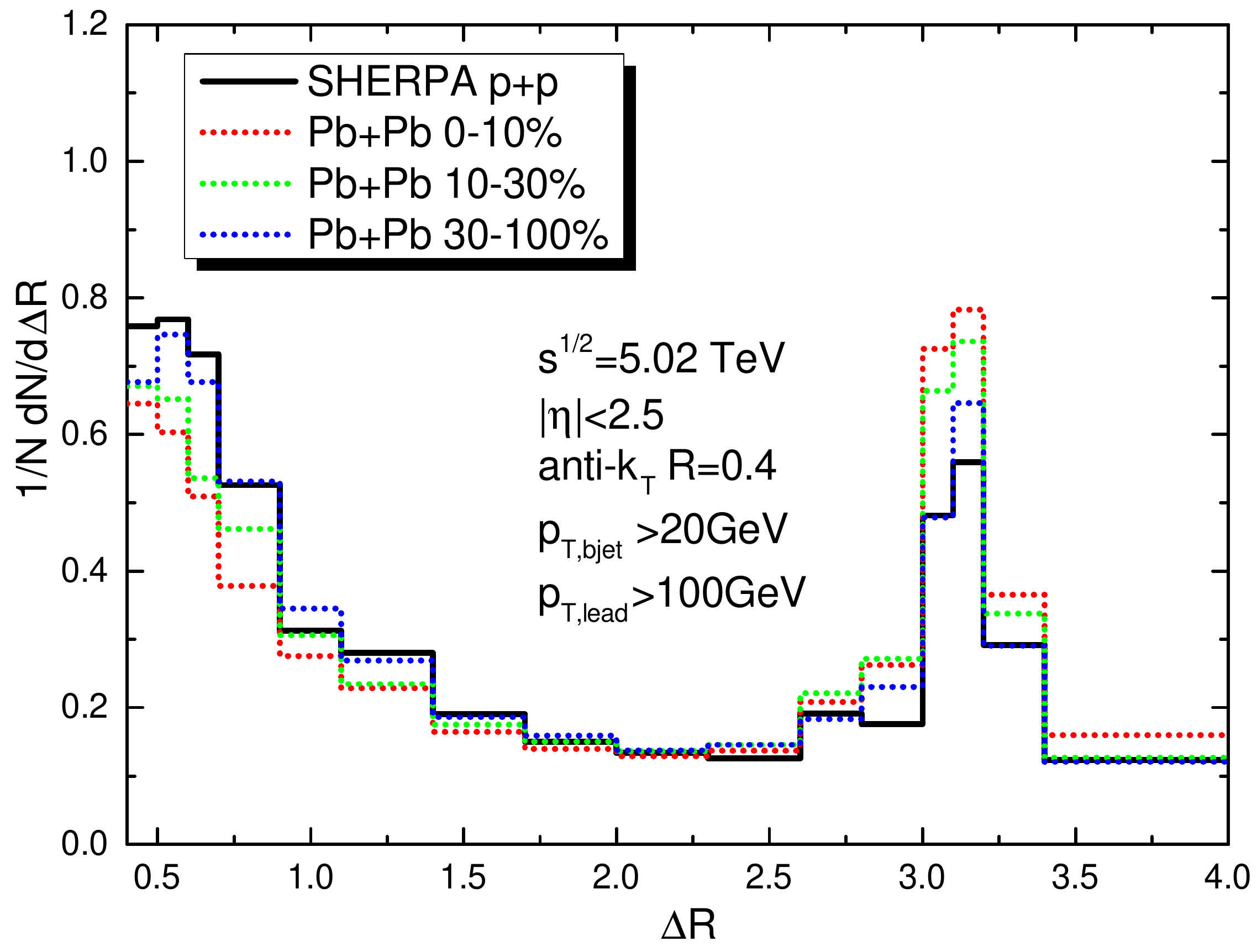, width=0.3\textwidth, clip=}}
 \subfigure[]{\label{fig:aveptphi}
  \epsfig{file=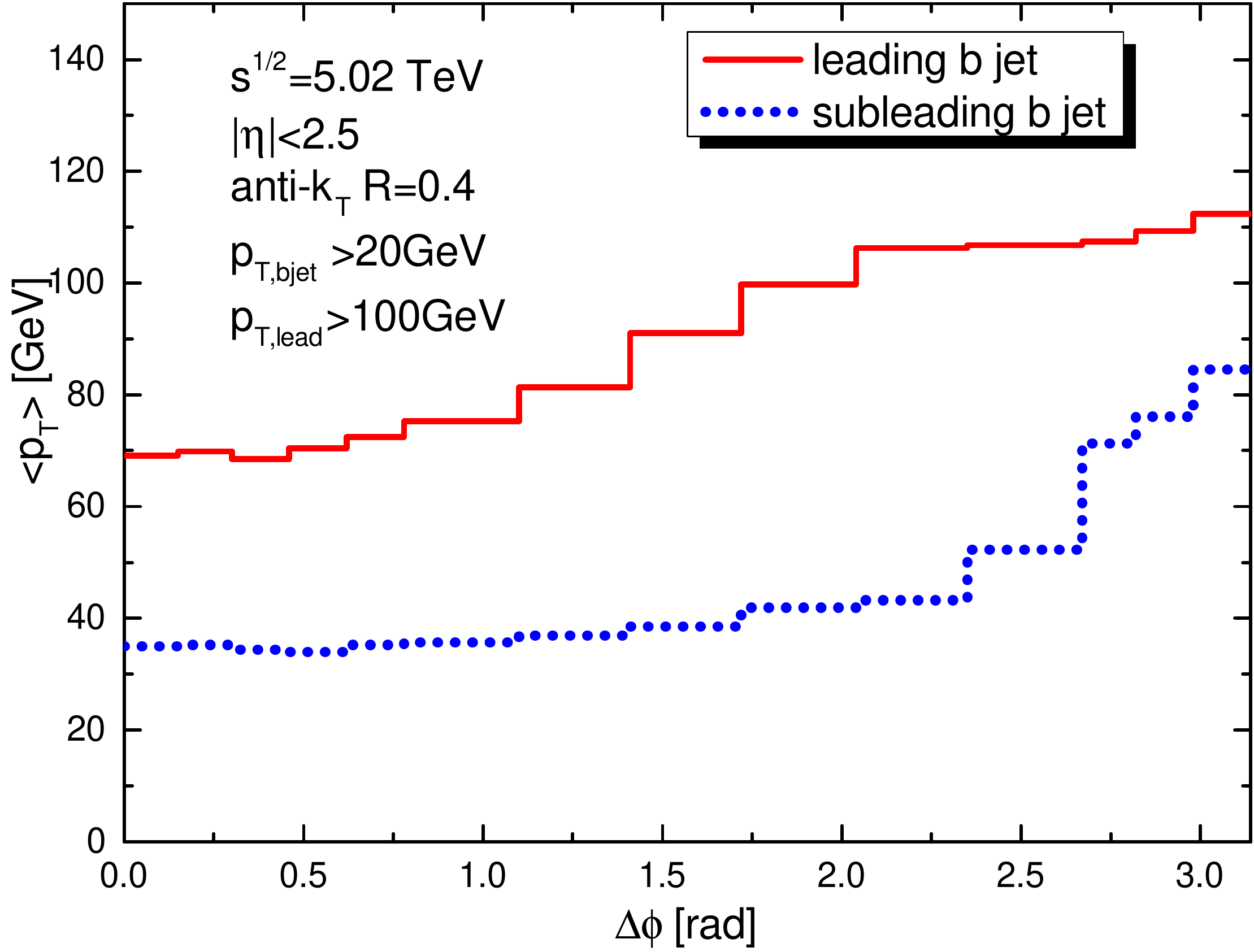, width=0.3\textwidth, clip=}}
  \caption{\scriptsize{(a) Normalized azimuthal angle distribution of $b\bar{b}$ dijet in p+p and Pb+Pb collisions at $\sqrt{s_{NN}}=5.02$~TeV. (b) Normalized angular distance distribution of $b\bar{b}$ dijet in p+p and Pb+Pb collisions at $\sqrt{s_{NN}}=5.02$~TeV. (c) The averaged $p_T$ distribution of leading and subleading b jet as a function of  their correlated azimuthal angle.}}
  \label{fig:correlation}
\end{figure}

Since the azimuthal angle distribution in small $\Delta \phi$ region is dominated by the gluon splitting processes while large $\Delta \phi$ region dominated by the flavour creation processes, it is useful to study the medium modification of the angular distribution between the two b jets. We present our predictions for the medium modification of azimuthal angle distribution of $b\bar{b}$ dijet in Pb+Pb collision in Fig.~\ref{fig:phi5020}. Our simulations show that the jet quenching effect would suppresses the peak in small angle region but also enhances another peak in large angle region in this normalized $\Delta \phi$ distribution. Actually, an overall suppression suffered on this distribution, but the stronger suppression in small angle region leads to a relative enhancement in larger angle region. We also give the calculated result of the medium modification for the angular distance distribution of $b\bar{b}$ dijet, shown in Fig.~\ref{fig:R5020}, where $\Delta R=\sqrt{(\Delta \eta)^2+(\Delta \phi)^2}$. A very similar trend as that in the $\Delta \phi$ distribution is observed: suppression in small $\Delta R$ region and enhancement in large $\Delta R$ region. To find out, we estimate the averaged transverse momentum distribution of the leading and subleading b jet in the $b\bar{b}$ dijet as a function of their correlated azimuthal angle shown in Fig.~\ref{fig:aveptphi}. We observe that the $\left\langle p_T \right\rangle$ in small $\Delta\phi$ region is lower than that in large $\Delta\phi$ region both for the leading and subleading b jet. It indicates that b jets produced by gluon splitting processes are ``softer'' than that produced by flavour creation processes. And then the in-medium energy loss would more easily shift the lower $p_T$ to a smaller value which is below the threshold of b-jet kinematic cut. This is the reason why stronger suppression suffered on small $\Delta \phi$ and $\Delta R$ region of $b\bar{b}$ dijet.


\section{Summary}
\label{sec:summary}
A Monte Carlo simulation which combines NLO+PS event generator SHERPA for pp baseline and Langevin transport equation including higher-twist gluon radiation to simultaneously describe the in-medium energy loss for both light and heavy partons has been implemented. We show the first theoretical results of transverse momentum imbalance for $b\bar{b}$ dijet in Pb+Pb collision and compare it with recent CMS data.  We find an increasing $p_T$ imbalance due to in-medium jet energy loss relative to the p+p reference which is consistent with the experiment data. Furthermore, we give the first prediction for the angular correlation of $b\bar{b}$ dijet in Pb+Pb collision at $\sqrt{s_{NN}}=5.02$~TeV. Since that the b jets produced in gluon splitting processes contain lower $p_T$ relative to that produced in flavour creation processes, the stronger suppression suffered on near side ($\Delta \phi\sim 0$) relative to that on away side ($\Delta \phi\sim\pi$) is predicted by our simulations.

\par This research is supported by the NSFC of China with Project Nos. 11435004, 11805167, and partly supported by China University of Geosciences (Wuhan) (No. 162301182691)


\begin{thebibliography}{99}
\bibitem{Wang:1991xy}
  X.~N.~Wang and M.~Gyulassy,
  Phys.\ Rev.\ Lett.\  {\bf 68} (1992) 1480; 
  M.~Gyulassy, I.~Vitev, X.~N.~Wang and B.~W.~Zhang,
  In *Hwa, R.C. (ed.) et al.: Quark gluon plasma* 123-191
  [nucl-th/0302077];
\bibitem{Norrbin:2000zc} 
  E.~Norrbin and T.~Sjostrand,
  Eur.\ Phys.\ J.\ C {\bf 17}, 137 (2000)

\bibitem{Dai:2018mhw}
  W.~Dai, S.~Wang, S.~L.~Zhang, B.~W.~Zhang and E.~Wang,
  arXiv:1806.06332 [nucl-th];

\bibitem{Zhang:2018urd} 
  S.~L.~Zhang, T.~Luo, X.~N.~Wang and B.~W.~Zhang,
  Phys.\ Rev.\ C {\bf 98}, 021901 (2018);

\bibitem{Sirunyan:2018jju}
  A.~M.~Sirunyan {\it et al.} [CMS Collaboration],
  JHEP {\bf 1803} (2018) 181
  [arXiv:1802.00707 [hep-ex]].
  
\bibitem{Gleisberg:2008ta}
  T.~Gleisberg, S.~Hoeche, F.~Krauss, M.~Schonherr, S.~Schumann, F.~Siegert and J.~Winter,
  JHEP {\bf 0902}, 007 (2009)
  [arXiv:0811.4622 [hep-ph]].

\bibitem{Ball:2014uwa}
  R.~D.~Ball {\it et al.} [NNPDF Collaboration],
  JHEP {\bf 1504}, 040 (2015)
  [arXiv:1410.8849 [hep-ph]].

\bibitem{Cacciari:2011ma}
  M.~Cacciari, G.~P.~Salam and G.~Soyez,
  Eur.\ Phys.\ J.\ C {\bf 72} (2012) 1896
  [arXiv:1111.6097 [hep-ph]].

\bibitem{Chatrchyan:2012dk}
  S.~Chatrchyan {\it et al.} [CMS Collaboration],
  JHEP {\bf 1204} (2012) 084
  [arXiv:1202.4617 [hep-ex]].

\bibitem{Aaboud:2016jed}
  M.~Aaboud {\it et al.} [ATLAS Collaboration],
  Eur.\ Phys.\ J.\ C {\bf 76} (2016) no.12,  670
  [arXiv:1607.08430 [hep-ex]].


\bibitem{Zhou:2016vwq}
  K.~Zhou, W.~Dai, N.~Xu and P.~Zhuang,
  Nucl.\ Phys.\ A {\bf 956} (2016) 120
  [arXiv:1601.00278 [hep-ph]].

\bibitem{Neufeld:2010xi} 
  R.~B.~Neufeld,
  Phys.\ Rev.\ D {\bf 83}, 065012 (2011)

\bibitem{Guo:2000nz}
  X.~F.~Guo and X.~N.~Wang,
  Phys.\ Rev.\ Lett.\  {\bf 85} (2000) 3591
  [hep-ph/0005044].

\bibitem{Zhang:2003yn} 
  B.~W.~Zhang and X.~N.~Wang,
  Nucl.\ Phys.\ A {\bf 720}, 429 (2003).

\bibitem{Zhang:2003wk}
  B.~W.~Zhang, E.~Wang and X.~N.~Wang,
  Phys.\ Rev.\ Lett.\  {\bf 93} (2004) 072301
  [nucl-th/0309040].
  
  

\bibitem{Majumder:2009ge}
  A.~Majumder,
  Phys.\ Rev.\ D {\bf 85} (2012) 014023
  [arXiv:0912.2987 [nucl-th]].
  
\bibitem{Cao:2017hhk} 
  S.~Cao, T.~Luo, G.~Y.~Qin and X.~N.~Wang,
  Phys.\ Lett.\ B {\bf 777}, 255 (2018).

\end{thebibliography}
\end{document}